\documentclass[%
 aip,
 jmp,%
 amsmath,amssymb,
reprint,
]{revtex4-1}

\usepackage[utf8]{inputenc}
\usepackage[english]{babel}
\usepackage{color}
\usepackage{graphicx}

\newcommand{\comments}[1]{} 				

\usepackage{dcolumn}
\usepackage{bm}

\begin{document}


\title{Current Induced Fingering Instability in Magnetic Domain Walls}

\author{J. Gorchon}%
\affiliation{Laboratoire de Physique des Solides, Université Paris-Sud, CNRS, UMR8502, 91405 Orsay, France}

\author{J. Curiale}
\affiliation{Laboratoire de Photonique et de Nanostructures, CNRS, UPR 20, 91460 Marcoussis, France}%
\affiliation{Laboratoire de Physique des Solides, Université Paris-Sud, CNRS, UMR8502, 91405 Orsay, France}
\affiliation{Consejo Nacional de Investigaciones Científicas y Técnicas, Centro Atómico Bariloche-Comision Nacional de Energía Atómica, Avenida Bustillo 9500, 8400 San Carlos de Bariloche, Río Negro, Argentina.}

\author{A. Cebers}%
\affiliation{University of Latvia, Zellu-8, Riga, LV-1002, Latvia}

\author{A. Lema\^{\i}tre}
\affiliation{Laboratoire de Photonique et de Nanostructures, CNRS, UPR 20, 91460 Marcoussis, France}%

\author{N. Vernier}
\affiliation{Institut d'électronique fondamentale, Université Paris-Sud, CNRS, UMR8622, 91405 Orsay, France}%

\author{M. Plapp}
\affiliation{Physique de la Matière Condensée, Ecole Polytechnique, CNRS, 91128 Palaiseau, France }%



\author{V. Jeudy}%
\email{vincent.jeudy@u-psud.fr}
\affiliation{Laboratoire de Physique des Solides, Université Paris-Sud, CNRS, UMR8502, 91405 Orsay, France}
\affiliation{Université Cergy-Pontoise, 95000 Cergy-Pontoise, France}

\date{\today}
\begin{abstract}

The shape instability of magnetic domain walls under current is investigated in a ferromagnetic (Ga,Mn)(As,P) film with perpendicular anisotropy. Domain wall motion is driven by the spin transfer torque mechanism. A current density gradient is found either to stabilize domains with walls perpendicular to current lines or to produce finger-like patterns, depending on the domain wall motion direction.  The instability mechanism is shown to result from the non-adiabatic contribution of the spin transfer torque mechanism.

\end{abstract}



\maketitle

%

Interface instabilities are encountered in a great variety of physical systems as liquids\cite{guyonBook01}, liquid-gas interfaces, ferro- and ferrimagnetic films\cite{hubertshafer00,seulPRA92,hagedorn70}, electrically polarizable and magnetic liquids\cite{tsori09,rosensweigJMMM83,cebersMagneto80}, intermediate state in type I superconductors\cite{prozorovNatPhys08,jeudyEPL06}... These instabilities originate from a competition between the surface tension which tends to favor flat interfaces and a destabilizing interaction as a gradient of external driving force (temperature, gravitational field, magnetic field...) or as long range dipolar interactions\cite{langerPRA92,JacksonPRE94} for quasi-two-dimensional systems\cite{seulscience95}. A crucial point for understanding interface dynamics as well as domain pattern formation is to determine the parameters controlling the instabilities and their formation mechanism.
 
In ferromagnetic systems, it was shown recently that domain walls (DWs) can be moved by a spin polarized current \cite{vernierEPL04,yamaguchiPRL04,mironNatMat11,curialePRL12} through the so-called spin transfer torque (STT) \cite{bergerPRB96,slonczewskiJMMM96,stilesPRB02,garatePRB09}. This has motivated an intense research effort for elucidating the physics of STT and for potential application in spin-electronics\cite{parkinScience08,grolier14}.  The STT acts as a driving force proportional to the current density. As expected by analogy with the well studied field-driven dynamics, essentially two dynamical regimes are observed. At low drive, DWs move in the pinning-dependent creep regime. Above a depinning threshold, the dynamics corresponds to flow regimes limited by dissipation\cite{curialePRL12}. Current-driven DW dynamics is most generally studied in narrow tracks, where DWs remain stable over the track width.
However, field and current-driven dynamics exhibit, in extended geometry, quite different behavior. A magnetic field acts essentially as a magnetic pressure pushing DWs with an average uniform velocity. In contrast, the current-driven creep regime was found to result in the formation of triangular domain-shapes\cite{moonPRL13}. In the flow regime \cite{vernierPRB13}, the DW velocity was shown to depend on the respective orientations of the DW and the current flow.
Those observations suggest a complex interplay between the DW shape and dynamics, and the STT magnitude. In this frame, it is particularly interesting to characterize the shape stability of DW driven by current. To address this issue, we investigated DW motion under current in wide geometries where instabilities induced by current and/or dipolar interactions can develop and be visualized. We used a (Ga,Mn)(As,P) thin film with perpendicular magnetization, as in this material, a wide range of dynamical regimes can be accessed thanks to its extraordinary weak current density required to induce DW motion. To get a better understanding of the role of current induced motion on DW stability, we introduce, on purpose, a progressive current density gradient by patterning our device in a semi-circular geometry.

In this letter, we show how the STT mechanism affects the domain pattern, and the DW shape stability. We found in particular that a current density gradient, depending on the DW motion direction, stabilizes or destabilizes the DW shape. A model, taking also into account surface tension and dipolar interactions, grasps the main features of DW stability.

\begin{figure}
\resizebox{0.98\columnwidth}{!}
{\includegraphics{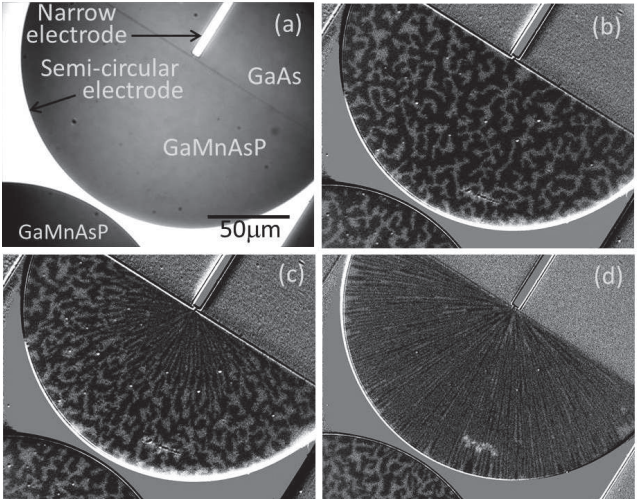}}
\caption{{\bf Current induced modification of magnetic domain pattern.} (a) Sample description. (b) Magnetic field driven domain pattern corresponding to the initial magnetic state. (c-d) Modification of domain pattern due to a DC current.  The current flows from the narrow to the semi-circular electrode ($j>0$) during 60~s. Its amplitude was $I=$~2.16~mA (image c) and I=2.98~mA (image d). The domain pattern is observed by magneto-optical Kerr microscopy. The two gray levels reflect the two opposite magnetization directions perpendicular to the (Ga,Mn)(As,P) film. $T=$~95~K.} 
\label{Fig1}       
\end{figure}

A 50~nm thick (Ga$_{0.95}$,Mn$_{0.05}$)(As$_{0.9}$,P$_{0.1}$) film was grown by low-temperature ($T=~250^\circ$C) molecular
beam epitaxy on a GaAs~(001) substrate \cite{lemaitreAPL08}. It was then annealed at $T= 250 ^\circ$C, for 1~h. Its magnetic anisotropy is perpendicular (saturation magnetization $M=~23\pm1$~kA/m) and its Curie temperature $T_c$ is 119$\pm$1~K. The semi-circular geometry (100 $\mu$m radius) was patterned by electron beam lithography and etching and then connected to a narrow (width $w=2\mu$m) electrode at the straight edge center and to a semi-circular electrode made of Ti (20~nm)/ Au (200~nm) layers (see Fig.~\ref{Fig1} (a)). The shape of magnetic domains and of DWs is controlled by differential polar magneto-optical Kerr microscopy with a 1~$\mu$m resolution in a cryostat with base temperature 95~K for all the experiments presented here (see Fig.~\ref{Fig1}(b-d)). The two gray levels correspond to opposite magnetization direction perpendicular to the film. Due to the semi-circular geometry, the electrical current lines are radial. The current density $j$ decays with the distance $r$ from the narrow electrode as $j(r) \approx I/\pi r h$ ($I$ is the injected current and $h=$~50~nm the film thickness) so that the gradient absolute value decreases progressively with $r$ as $\left|dj/dr\right|=\left|I \right|/(\pi h r^2)$. In the following by convention, $I>0$ (i.e. $j>0$) corresponds to a current flow form the narrow to the semi-circular electrode.

\begin{figure}
\resizebox{0.98\columnwidth}{!}
{\includegraphics{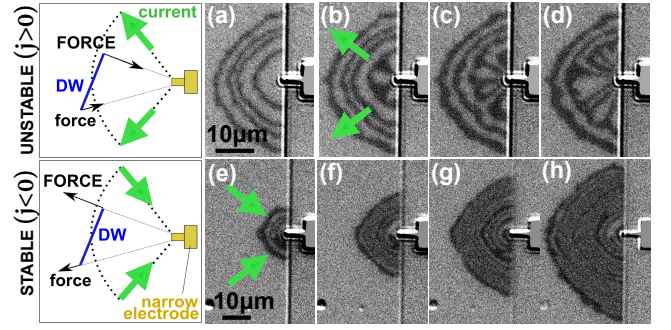}}
\caption{{\bf Instability of magnetic DW produced by a gradient of current density.} (Left frames) Stability of a domain wall (dotted arcs) placed perpendicularly to a current density gradient. A small tilt of an elementary wall length (blue segments) produces an asymmetry of the forces due to spin transfer (thin black arrows). A current flow (thick green arrows) in the direction of the narrow electrode ($j>0$, top frame) tends to destabilize the initial orientation while it tends to be stabilized for $j<0$ (bottom frame). 
(a-d) DW shape instability for $j>0$. (a) Initial state. (b-d) A 60~seconds DC current flow produces a finger  growth towards the narrow electrode. Increasing the current magnitude ($I=$~0.70; 1.10; 1.20~mA for image (b), (c), (d), respectively) enhances the distance at which semi-circular DWs become unstable. 
(e-h) Stable radial DW growth for $j<0$. The sample initially in an homogeneous magnetic state is submitted to a current pulse of amplitude -2.164~mA of increasing duration (10~$\mu$s; 100~$\mu$s; 1~ms and 10~ms  for images (e), (f), (g) and (h), respectively). The propagation front remains almost semi-circular. $T=$~95~K.}
\label{Fig2}       
\end{figure}

First evidences of domain wall shape instability are shown in Figure \ref{Fig2}. A set of semi-concentric magnetic domains centered on the narrow electrode (see Figs. \ref{Fig2} (a) and (e)) was prepared) using current induced stochastic domain nucleation and DW propagation (see ref. \onlinecite{gorchonPRL14} for details) starting from a uniform magnetization state. Next a DC current was injected between the two electrodes for a fixed duration after which an image was acquired. The sequence is repeated for Fig.~\ref{Fig2} (b-c) with increasing current (during 60~s at $I=0.7$, 1.1 and 1.2~mA) and for Fig.~\ref{Fig2}(f-h) with increasing duration (100~$\mu$s, 1~ms and 10~ms at $I=-2.164$~mA). 
The DW motion observed in Figs. \ref{Fig2}(b-c) and (f-h) originates from the spin transfer torque. In ferromagnets, the electrical current is spin-polarized and carriers crossing a DW exert a torque on the local magnetic moment, which results on DW propagation. In (Ga,Mn)(As,P) films with perpendicular anisotropy, DW motion is in the opposite direction to the current\cite{curialePRL12},as it can be observed. In this experiment, different DW dynamical regimes are expected to occur due to the decay of $j$ with $r$. Close to the narrow electrode, the current density ($j \approx I/hw=$~10-20~GA/m$^2$, where $w=$~2~$\mu$m is the width of the narrow electrode) is sufficiently large for the flow regime to be reached\cite{curialePRL12} while pinning dependent regimes are expected to occur in the other parts of the device.
%

The most original in the results shown in Fig.~\ref{Fig2} is dependance of the shape of domains on the current polarity. For $j>0$, the semi-circular symmetry of domains breaks. In Fig.~\ref{Fig2}(b), the black domain next the narrow electrode expands toward the electrode by forming finger-like shapes. As the current amplitude increases (Figs.~\ref{Fig2}(c) and (d)), the instability process takes also place in domains located farther away from the electrode. On the opposite, for $j<0$, the semi-circular geometry is conserved. The shape of domain walls is stable during the motion.
%
%
We explain now, first qualitatively, the contribution of the current density gradient to the domain wall stability.
The left frames of Fig.~\ref{Fig2} give a schematic description of this mechanism. Let us consider a slightly tilted elementary DW segment. Due to the current density gradient, the two segment ends experience a different STT amplitude. It is larger for the one closer to the narrow electrode. This asymmetry is the driving mechanism for the DW stabilization or destabilization. When the $j$ is negative, the STT force points away from the narrow electrode and the DW segment moves away from the electrode. However, the lagging segment extremity experiences a stronger STT force than the opposite end, therefore acting as a restoring force. The DW remains stable during its motion. In turn, this mechanism is responsible for the DW destabilization when $j>0$ (opposite DW motion direction) since the STT force is stronger for the forward end. It eventually leads to domain growth along the current lines. This behavior shares similarities with the Rayleigh-Taylor instability\cite{guyonBook01}, when a heavier liquid is above a lighter one.   

This instability mechanism has dramatic consequences on domain pattern formation up to very large radii and hence very low DW velocities (see Figs. \ref{Fig1}(b-d)). 
Fig.~\ref{Fig1}(b) shows an initial demagnetized state (obtained before applying any magnetic field or current). The magnetic domains with opposite magnetization direction present a self-organized pattern, as usually observed in ferromagnetic films with perpendicular anisotropy. The typical domain width and spacing ($\approx$ 10~$\mu$m and $\approx$ 20~$\mu$m, respectively) results from a balance between the positive DW energy and long-range magnetic interactions between domains \cite{gourdonPRB07}. The domain shape corresponds to randomly oriented corrugated lamellae.

%
After applying a positive DC current ($j>0$) during 60~s (see Figs. \ref{Fig1}(c-d)), domains tend to be aligned radially. For the largest current value ($I=2.98$~mA), the domain pattern is modified over the full sample surface area, as observed in Fig. \ref{Fig1}(d). The DW are aligned along the current lines, a consequence of the gradient induced destabilization mechanism described earlier. 
%
%
We can get an insight of the DW organization dynamics when injecting lower current values. In that case, the  domain pattern modification remains spatially limited by a semi-circular boundary centered on the narrow electrode as seen in Fig. \ref{Fig1} (b) ($I=2.16$ ~mA). 
%
%
Indeed, sufficiently far from the narrow electrode, DWs follow dynamical regimes controlled by DW pinning and thermal activation. In those regimes, the DW velocity varies exponentially with the driving force. As the STT amplitude decreases as $\left|I\right|/r$, the DW velocity considerably reduces as it is located at a greater distance from the high current density regions close to the narrow electrode. Therefore, for a limited current pulse duration (60~s), each given current value $I$  defines a semi-circular clear-cut boundary separating regions with unmodified patterns (at the scale of the experimental spatial resolution $\approx$~1~$\mu$m) from regions presenting significant DW displacements, as observed in Fig.~\ref{Fig1}(c). 

%
%
%

%
%
 
\begin{figure}
\resizebox{0.98\columnwidth}{!}
{\includegraphics{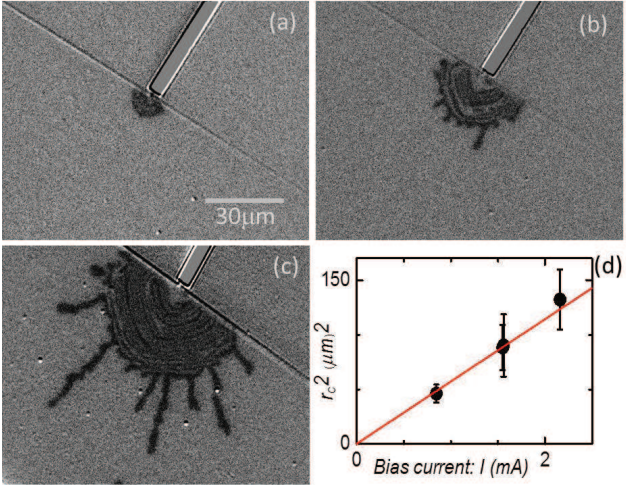}}
\caption{{\bf Instability of magnetic domain wall produced by dipolar interaction.} The images were obtained for a constant current ($I=$~-1.55~mA) directed towards the narrow electrode ($j<0$) and different durations ((a):10~ms; (b): 690~ms; (c): 29.7~s). $T= $ 95~K. (d) Square of the critical instability radius as a function of the bias current. The line corresponds to the best fit of the theoretical prediction.}
\label{Fig4_5}       
\end{figure}

At this point, we have shown how a current density gradient can stabilize or destabilize a DW. However, we have not considered yet, how this mechanism competes or cooperates with the other mechanisms involved in  DW stability, as dipolar interactions and the DW surface tension. To that end, we extended the experiment described in Fig.~\ref{Fig2}(e-h) ($I=-1.55$~mA), to longer current pulses. As previously, the sample was first prepared in a fully homogeneous magnetized state.  Negative current pulses were injected with 10~ms, 690~ms and 29.7~s durations. In this situation, the gradient acts as a stabilization contribution. For the shortest duration, the domains present a semi-circular shape (see Fig.~\ref{Fig4_5}(a)) that reflects the current line symmetry as already observed in Figs.~ \ref{Fig2}(e-h). However, for the longest durations (see Fig. \ref{Fig4_5}(b-c)), the semi-circular shape of the domains with the largest radius becomes unstable and finger-like domain growth is observed.  The  finger width is close to typical size of domain patterns observed in the demagnetized configuration (see Fig. \ref{Fig1}(b)). This behavior strongly points toward the dipolar interactions as the destabilization mechanism.
The critical instability radius $r_c$ at which finger-shaped domains start to grow was measured systematically as a function of the injected current $I$. As reported in Fig.~\ref{Fig4_5}(d), $r_c^2$ is found to vary linearly with $I$, i.e., the critical radius is associated to a well defined critical current density gradient ($\left|dj/dr\right|)_c=\left|I\right|/(\pi h r_c^2)$). 
Therefore, the DW shape instability observed in Fig. \ref{Fig4_5}(a-c) occurs when the current density gradient becomes too weak to stabilize the DWs perpendicular to current lines against the dipolar interactions.

To get a more quantitative insight on the DW shape instabilities, we have elaborated a model, which describes the stability of a flat DW subjected to an electrical current gradient. %
%
%
The model considers a ferromagnetic layer of thickness $h$ along to the $z$-direction and a flat DW, aligned along the $x-z$-plane, which separates two domains with opposite magnetization directions. The DW is submitted to a current flow exhibiting a gradient in the $y$-direction. The magnetization vector is given as $\overrightarrow{M}=M(\sin\theta\cos\varphi,\sin\theta\sin\varphi,\cos\theta)$. In the perturbed state, the DW position is given by the equation $y=q(x,t)$.   
%
%
The DW shape stability analysis  is based on the Landau-Lifshitz-Gilbert equation and follows the calculation of Refs. \onlinecite{thiavilleEPL05} and \onlinecite{malozemoffBook79}. The full calculation is detailed in the supplemental material\cite{supp_info}. For a weakly perturbed DW, the equations of motion are:
\begin{equation}
\label{equation_stab1}
\gamma\left(\mu_0 M+\frac{2\psi_{2M}(q,h)}{h}\right)-\frac{2A\gamma}{M\Delta}\frac{\partial^2q}{\partial x^2}=\dot{\varphi}-\frac{\alpha\dot{q}}{\Delta}+\frac{\beta u}{\Delta}\\
\end{equation}

and
\begin{equation}
\label{equation_stab2}
\frac{\mu_0}{2}\gamma M \sin2\varphi-\frac{2A\gamma}{M}\frac{\partial^2\varphi}{\partial x^2}=-\alpha \dot{\varphi}+\frac{u}{\Delta}-\frac{\dot{q}}{\Delta},
\end{equation}
where $\gamma$, $\alpha$ $\beta$  are the gyromagnetic factor, the Gilbert damping parameter and so-called non-adiabatic term, respectively. $\Delta=\sqrt{A/K}$ is the domain wall thickness parameter, where $A$ and $K$ are the spin stiffness and the anisotropy constant, respectively. The parameter $u$ is the spin drift velocity defined by $u = \frac{jP_c g\mu_B}{2eM}$, where $j$, $P_c$, $g$, $\mu_B$, and $e$ ($<$0), are the current density, the current spin polarization, the Land\'{e} factor, the Bohr magneton, and the electron charge, respectively.
In Eq.~\ref{equation_stab1}, $\psi_{2M}(q,h)$ is a potential describing the dipolar interaction between the DW magnetization and the field created by the two magnetic domains with opposite magnetization.

For a small perturbation $\delta \varphi$, $\delta q$ of the DW, the perturbation of the spin drift velocity can be written $\delta u \approx (du/dq)\delta q$. Assuming a steady DW motion ($\dot{\varphi}=0$) and looking for solutions of the type $\delta q \sim \delta q_0 exp(ikx)$, Eq.~\ref{equation_stab1} reads

\begin{equation}
\label{equation_stab_perturb_1}
\left[F +\frac{du}{dq}\frac{\beta}{\mu_0 M} \frac{h^2}{\gamma \Lambda^2} \right] \delta q_0= \frac{\alpha h^2}{\mu_0 M \gamma \Lambda^2}\frac{d(\delta q_0)}{dt}
\end{equation}
where we have introduced the exchange length $\Lambda$ defined by $A=\mu_0 M^2 \Lambda^2/2$, the magnetic Bond number\cite{characteristic_length} $B_m=\mu_0 (2M)^{2}h/(4 \pi\sigma)$  with the DW surface energy given by $\sigma =4\sqrt{AK}$. In Eq.~\ref{equation_stab_perturb_1}, the function $F$ is given by $F= 4B_m(\gamma_E + \log{(kh/2)}+K_0(kh))-(kh)^2$, where $K_0(kh)$ is the McDonald function and $\gamma_E$ ($=$0.5772) the Euler constant. 

The differential equation Eq.~\ref{equation_stab_perturb_1} shows that a flat DW is unstable if the coefficient in brackets on the left hand side is positive. The instability thus results from a competition between the dipolar energy (the first terms of function $F$), the DW surface tension (the term $(kh)^2$ in $F$) and the STT gradient  ($\propto du/dq$ in Eq.~\ref{equation_stab_perturb_1}). One should note that only the non-adiabiatic contribution ($\propto \beta$) of the STT plays a role in DW stability.\cite{supp_info} The fastest instability growth rate corresponds to the function $F$ maximum which is equal to $F_{max}= 2B_m \exp{(1-2\gamma_E-2/B_m)}$ and to a wavelength $\lambda$ $=\pi h \exp{ \left(\gamma_E +1/B_m-1/2\right)}$, in the limit of small $kh$.

For the semi-circular geometry considered in the letter, the conservation of the current $I= j\pi rh$ leads to $\frac{du}{dq}=-\frac{IP_cg\mu_B}{ \pi h r^2 2eM}$. For a current flow from the narrow electrode ($j>0$, i.e., $du/dq>0$), the flat DW is always unstable. This corresponds to the case presented in the top frames of  Fig.~\ref{Fig2} for which both the current density gradient and the dipolar interactions have a destabilizating contribution.  For the opposite current direction ($j<0$, i.e., $du/dq<0$), DW instability occurs below a gradient threshold corresponding to a critical radius given by $r_c^2=I \frac{C}{F_{max}}$, where $C=\frac{h \beta P_c g \mu_B}{4 \pi A \gamma e}$. Above this critical radius, the stabilization contribution of the gradient becomes too weak to counteract the effect of dipolar interactions. 

Comparing those predictions to  the experimental results requires the evaluation of the magnetic Bond number $B_m$. 
First, $B_m$ can be estimated from the critical radius $r_c$, measured in Fig.~\ref{Fig4_5}(d). The data best fit  gives a ratio $r_c^2/I=C/F_{max}=$~58$\pm$3~$\mu$m$^2$/mA. Assuming $\beta=$~0.3,\cite{curialePRL12} $P_c=$~0.5, $g=$~2, $\mu_B$=~9.3 10$^{-24}$J.T$^{-1}$, $\gamma=$ 1.76 $ 10^{11}$Hz.T$^{-1}$ and $A=$~0.07$\pm$0.03~pJ/m,\cite{haghgooPRB10} we have $1/F_{max}\approx$~10000 and $B_m\approx$~0.25.\cite{variability}
$B_m$ can also be deduced from the number $n$ of fingers observed in Fig. \ref{Fig4_5} (b) and (c). Indeed, assuming $n$ to remain constant after the onset of the DW instability (occurring for $r=r_c$), the critical perturbation wavelength reads $\lambda=\pi r_c/n$ whose value was extracted from a statistical analysis, $\lambda=$~4$\pm$1$\mu$m. The prediction for $\lambda$ leads to $B_m=$~0.34$\pm$0.04, a value close to the previous estimation.
%
Finally, $B_m$ can also be estimated independently from micromagnetic parameters (see ref.~\onlinecite{haghgooPRB10}) since $B_m=\mu_0 (2M)^{2}h/(4 \pi\sigma)$ with $\sigma=4 \sqrt{AK}$. The obtained Bond number equals 0.3$\pm$0.1 and presents a good quantitative agreement with the two previous estimations. This unambiguously demonstrates that the domain wall fingering instability, observed for $j<0$, originates from a competition between the dipolar interactions and the effect of the current gradient whose magnitude is shown to be proportionnal to non-adiabatic contribution of the STT.

%

In conclusion, these results show that the domain wall orientation with respect to a current flow is very sensitive to current density gradients in current induced DW motion experiments. They unveil some potential weaknesses for future devices relying on complex circuits where these gradients are ubiquitous. Yet, they also give us some interesting directions to propagate and manipulate DW over large surface, by taking advantage of the gradient controlled stability.

{\bf Acknowledgements}
The authors wish to thank J. Miltat for his careful reading of the manuscript. 
This work was partly supported by the French projects DIM C'Nano IdF (Région Ile-de-France), ANR-MANGAS (No. 2010-BLANC-0424), RTRA Triangle de la physique Grants No. 2010-033TSeMicMagII and No. 2012-016T InStrucMag and the LabEx NanoSaclay, and by the Argentinian project PICT 2012-2995 from ANPCyT and UNCuyo Grant No. 06/C427. This work was partly supported by the french RENATECH network.


\begin{thebibliography}{}
\bibitem{characteristic_length} The characteristic length of bubble materials $l=\sigma/(\mu_0M^2)$ is related to the Bond number $B_m$ by $l \pi B_m=h$.
\bibitem{variability} Note that this estimation is weakly dependent on the values of $\beta$ and $P_c$: a variation of the product $\beta P_c$ by a factor $\pm$2 changes $B_m$ by less than 10\%.
\end{thebibliography}
\end{document}